\documentclass[prl, twocolumn,showpacs,preprintnumbers,amsmath,amssymb,superscriptaddress]{revtex4-1}

\usepackage{graphicx}
\usepackage{dcolumn}
\usepackage{bm}
\usepackage{epsf}

\begin{document}

\newcommand{\Jm}{J_{\mathrm{max}}}
\newcommand{\lm}{\lambda_{\mathrm{max}}}

\title{Largest Lyapunov exponents for lattices of interacting classical spins}

\author{A. S. de Wijn}
\email{dewijn@fysik.su.se}
\affiliation{Department of Physics, Stockholm University, 106 91 Stockholm, Sweden}
\affiliation{Institute for Molecules and Materials, University of Nijmegen, P.O Box 9010, 6500 GL  Nijmegen, The Netherlands}
\author{B. Hess}
\affiliation{Institute for Theoretical Physics, University of Heidelberg, Philosophenweg 19, 69120 Heidelberg, Germany}
\author{B. V. Fine}
\email{B.Fine@thphys.uni-heidelberg.de}
\affiliation{Institute for Theoretical Physics, University of Heidelberg, Philosophenweg 19, 69120 Heidelberg, Germany}

\date{\today}

\begin{abstract}
We investigate how generic the onset of chaos in interacting many-body classical systems is in the context of lattices of classical spins with nearest-neighbor anisotropic couplings.
Seven large lattices in different spatial dimensions were considered.
For each lattice, more than 2000 largest Lyapunov exponents for randomly sampled Hamiltonians were numerically computed.
Our results strongly suggest the absence of integrable nearest-neighbor Hamiltonians for the infinite lattices except for the trivial Ising case.
In the vicinity of the Ising case, the largest Lyapunov exponents exhibit a power-law growth, while further away they become rather weakly sensitive to the Hamiltonian anisotropy.
We also provide an analytical derivation of these results.

\end{abstract}

\maketitle

The concept of microscopic chaos in many-particle systems plays an essential role in the foundations of statistical physics~\cite{Gibbs-02,Krylov-79,Gaspard-98}.
In classical systems, chaos is defined by the appearance of exponential instabilities with respect to infinitesimal perturbations of the initial conditions.
The spectrum of these instabilities is characterized by a set of eigenvalues -- Lyapunov exponents -- and the corresponding eigenvectors.
In general, interacting, many-particle, classical systems are expected to be chaotic.
The largest Lyapunov exponents and the whole Lyapunov spectra have been calculated numerically for classical many-body systems, such as gases of hard-core particles~\cite{posch1,posch2}, fluids with soft interactions~\cite{soft}, and lattice two-dimensional rotators~\cite{soft,ruffo}, and analytically in a few cases~\cite{prlramses,lagedichtheid,astridhenk,mareschal,cylinders,roc_astrid_henk,barnett}.
Nevertheless, little is known about the universality of the features of the Lyapunov spectra, in particular when it comes to the systems with smooth dynamics.
It is also not clear how generic the onset of chaos actually is, and what happens in many-particle systems in the vicinity of integrable, i.e. nonchaotic, limits of the microscopic Hamiltonians.

This Letter deals with the above issues on the basis of a systematic numerical and analytical investigation of the systems of interacting classical spins. These systems have been extensively studied, e.g., in the context of the spin diffusion problem~\cite{Muller-88,Gerling-90,Bonfim-92}, but their Lyapunov spectra have not yet been computed.
Finite classical spin systems are also known to exhibit nontrivial integrable limits~\cite{Steinigeweg-09}. 
Another aspect of our motivation is 
the connection to the quantum case. There is a growing appreciation that generic interacting many-particle quantum systems exhibit relaxation behavior similar to their classical chaotic counterparts, in particular as far as nuclear spin decays in solids are concerned~\cite{Fine-04,Fine-03,Fine-05,Elsayed-11,Morgan-08,Sorte-11,Sorte-12,Meier-12,Pastawski-00}.
In a broader context, 
the important conceptual development in the present study is the massive character of the numerical investigation covering an entire class of Hamiltonians.  

In this Letter, we consider several large lattices of interacting classical spins.
For each lattice, we compute the largest Lyapunov exponent $\lm$ for several thousand randomly selected microscopic Hamiltonians.
Thereby, we obtain the dependence of the largest Lyapunov exponent on the anisotropy of the spin-spin interaction.
Our results give a strong indication of the absence of integrable Hamiltonians for infinite spin lattices with nearest-neighbor interaction besides the trivial Ising case (explained below).
We also find that, to the extent afforded by our numerical accuracy, the system becomes chaotic in the immediate vicinity of the integrable Ising case, with $\lm$ exhibiting a universal power-law scaling.
Further away from the Ising limit, $\lm$ becomes only weakly dependent on the Hamiltonian anisotropy, especially for bipartite lattices.
The letter is concluded with a simple analytical derivation that describes the above-mentioned numerical results.

\begin{figure} \setlength{\unitlength}{0.1cm}

\begin{picture}(100, 70)
{ 
\put(0, 0){ \epsfxsize= 3.2in \epsfbox{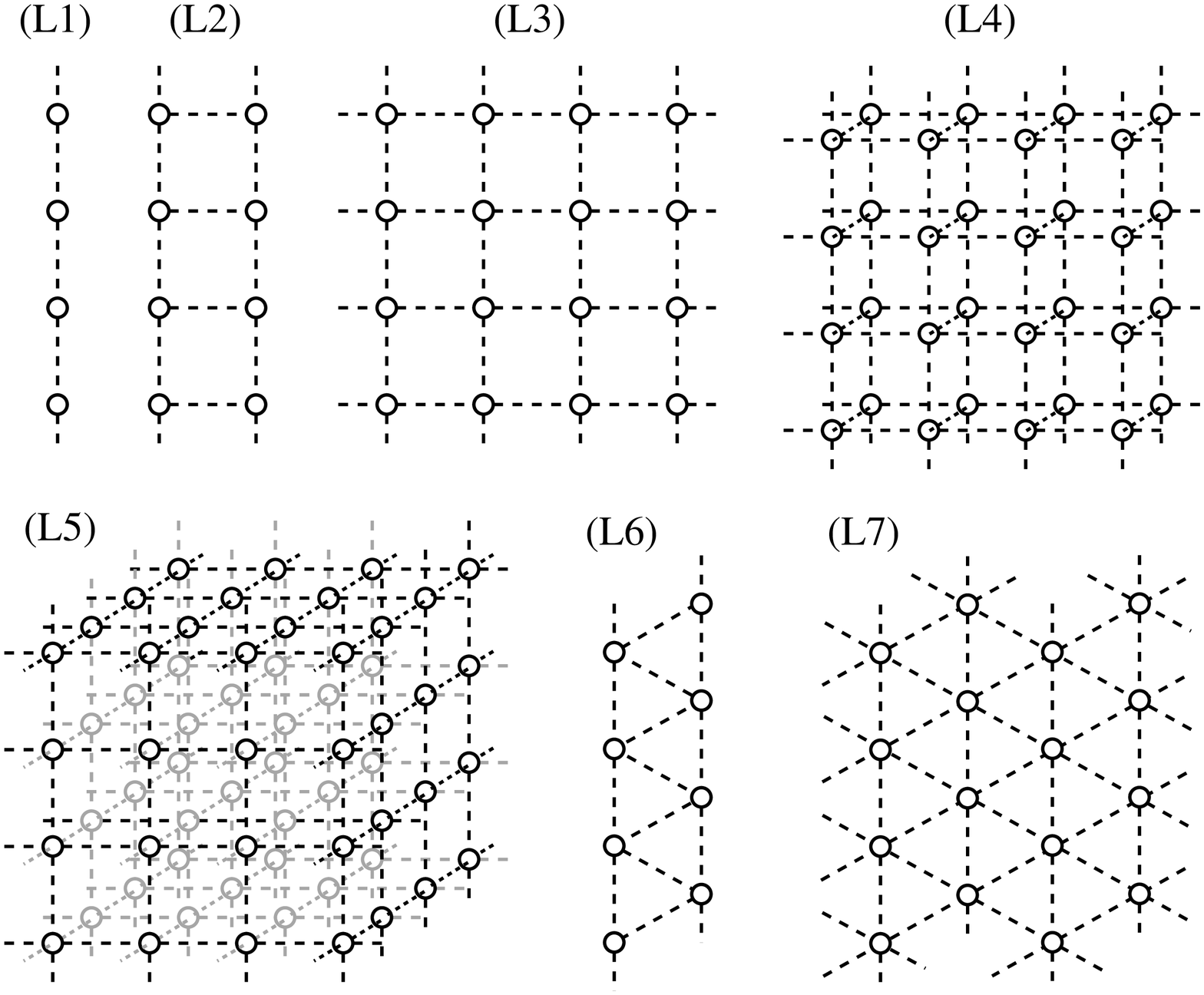} }
}
\end{picture}
\caption{Lattices investigated in this work.  Bipartite: (L1) chain, (L2) rectangular ladder, (L3) square lattice, (L4) bilayer of square lattices, (L5) cubic lattice.
Non-bipartite: (L6) triangular ladder, (L7) triangular lattice. 
}
\label{fig:lattices} 
\end{figure}

We investigate the seven lattices shown in Fig.~\ref{fig:lattices}: (L1) a chain, (L2) a rectangular ladder, (L3) a square lattice, (L4) a bilayer of square lattices, (L5) a cubic lattice, (L6) a triangular ladder and (L7) a triangular lattice. 
The interaction Hamiltonian for each lattice is of the nearest-neighbor (NN) type with periodic boundary conditions:
\begin{equation}
H=\sum_{i<j}^{\hbox{\scriptsize NN}} J_x S_{ix} S_{jx}+J_y S_{iy} S_{jy}+J_z S_{iz} S_{jz}~,
\label{H}
\end{equation}
where $(S_{ix}, S_{iy}, S_{iz}) \equiv {\mathbf S}_i$ are the three projections of the $i$th classical spin normalized by the condition ${\mathbf S}_i^2 = 1$.

We numerically integrate the equations of motion associated with the Hamiltonian~(\ref{H}): $\dot{\mathbf S}_i = {\mathbf S}_i \times {\mathbf h}_i$~\cite{footnotespins}, where ${\mathbf h}_i$ is the local field given by the expression
\begin{equation}
\mathbf h_i = \sum_{j(i)} J_x S_{jx} {\mathbf e}_x + J_y S_{jy} {\mathbf e}_y + J_z S_{jz} {\mathbf e}_z~.
\label{eom}
\end{equation}
Here ${\mathbf e}_x$, ${\mathbf e}_y$ and ${\mathbf e}_z$ are the unit vectors along the respective directions, and $j(i)$ implies the summation over the nearest neighbors of the $i$-th lattice site.
We use the fourth-order Runge-Kutta algorithm and choose a time step of $0.005$, sufficiently small so that, on the time scales of our simulations, energy is conserved.
The initial conditions of each trajectory are chosen randomly on the energy shell with zero total energy, which corresponds to infinite temperature~\cite{footnoteinfinite}.

In order to obtain $\lm$, we numerically calculate a phase space trajectory and, at every time step, track the evolution of a tangent space vector that defines an infinitesimal perturbation of it~\cite{Benettin-80,Elsayed-11}. The asymptotic average growth rate of this vector is equal to the largest Lyapunov exponent.
In all cases, $\lm$ becomes size-independent for sufficiently large lattices.

For each lattice, we computed $\lm$ for many combinations of the coupling constants randomly selected on the ``interaction sphere'' $J_x^2 + J_y^2 + J_z^2 = 1$. Specifically, 8000 combinations selected from the isotropic distribution were selected for the lattices (L1), (L3), (L5), and 2000 combinations for each of the remaining lattices.  
In addition, in order to investigate the scaling behavior near the Ising limit, we have processed more combinations in the vicinity of $J_z=1$: 1000 for the lattices (L1), (L3), and (L5) and 250 for the others.

\begin{figure}[t] \setlength{\unitlength}{0.1cm}
\begin{picture}(100, 110)
{
\put(2,105){(a)} 
\put(0,110){ \epsfxsize= 2.3in \rotatebox{270}{\epsfbox{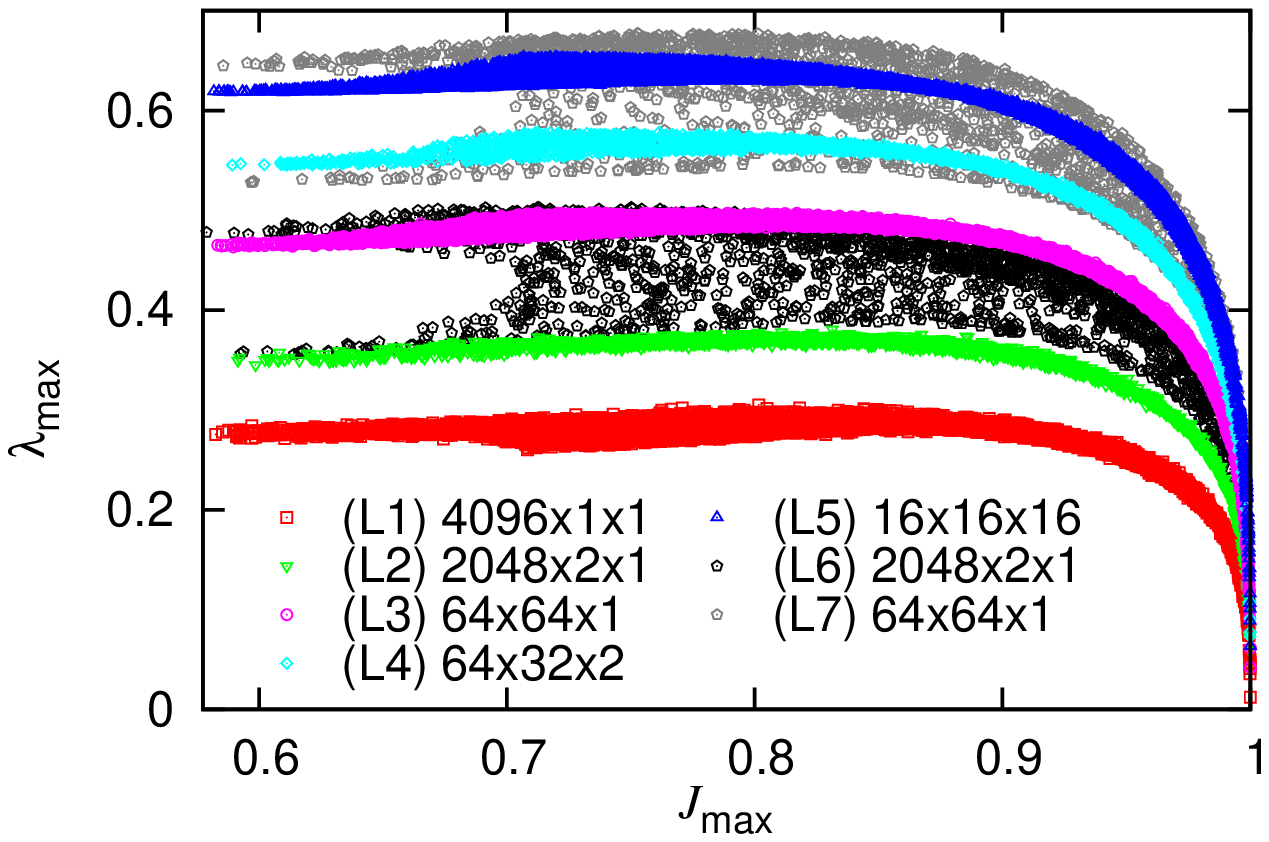}} }
\put(2,50){(b)} 
\put(0,55){ \epsfxsize= 2.3in \rotatebox{270}{\epsfbox{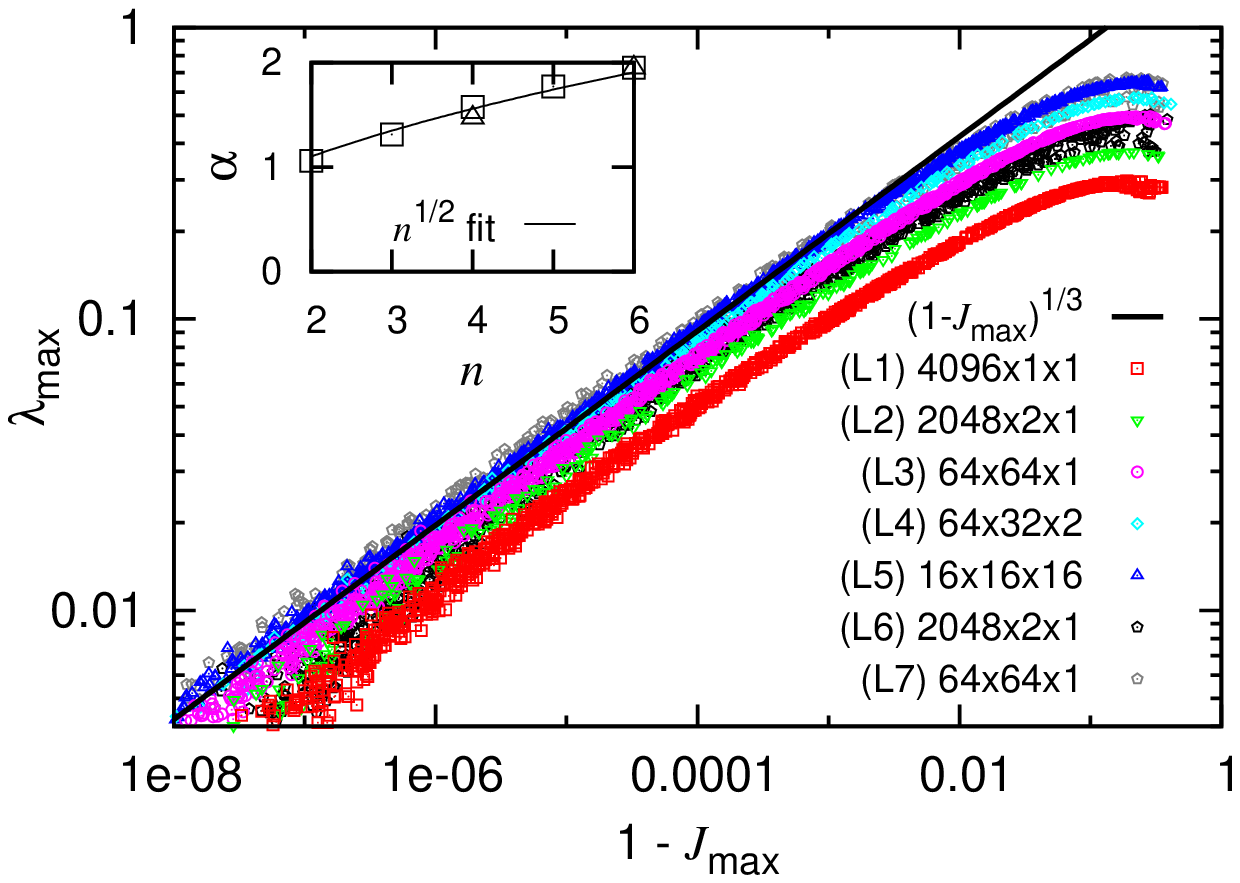}} }
}
\end{picture}
\caption{(Color online) Largest Lyapunov exponents. Each point represents one $\lm$ obtained numerically for a lattice indicated in the plot legend with one randomly chosen set of values $J_x$, $J_y$ and $J_z$ as described in the text (a) Linear plot. (b) Log-log plot.
The inset shows the prefactor $\alpha$ of the power-law fit $\lm = \alpha (1-\Jm)^{1/3}$ as a function of the number of nearest neighbors $n$ with squares for (L1-L5), triangles for (L6,L7) and solid line for the fit $\alpha \cong n^{1/2}$.
}
\label{fig:all} 
\end{figure}

\begin{figure}[t] \setlength{\unitlength}{0.1cm}
\begin{picture}(100, 55)
{
\put(0,55){ \epsfxsize= 2.3in \rotatebox{270}{\epsfbox{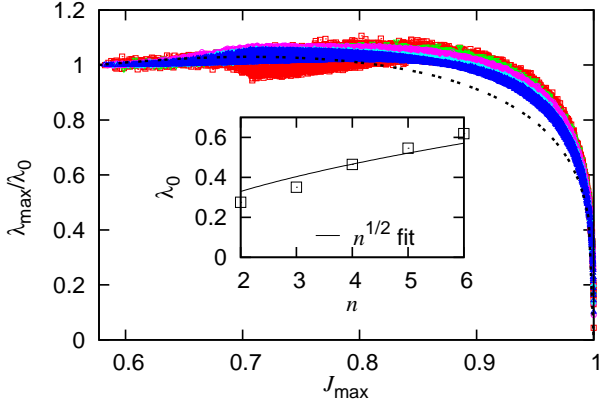}} }
}
\end{picture}
\caption{(Color online) Largest Lyapunov exponents for the bipartite lattices (L1-L5) from Fig.~\ref{fig:all}(a) rescaled by dividing by $\lambda_0 \equiv \lm(1/\sqrt{3})$. The dotted line represents $\lm \cong  \Jm^{1/2} (1-\Jm^2)^{1/4}$. The inset shows $\lambda_0$ as a function of the number of nearest neighbors $n$.
}
\label{fig:rescaling} 
\end{figure}

Our main numerical findings, namely
$\lm$ as a function of the parameter $\Jm \equiv \hbox{max}(|J_x|, |J_y|, |J_z|)$,
are presented in Fig.~\ref{fig:all}.
The maximum value $\Jm=1$ is realizable only in the Ising case, and thus represents the integrable limit with $\lm = 0$.
The minimum value of $\Jm$ is $1/\sqrt{3}$.
It corresponds to either Heisenberg case $J_x = J_y = J_z$ or ``anti-Heisenberg'' case $J_x = J_y = - J_z$ (or equivalent cases).
In Fig.~\ref{fig:rescaling}, we present the rescaled plots $\lm(\Jm)/\lm(1/\sqrt{3})$ for lattices (L1-L5). 
On the basis of the results presented in Figs.~\ref{fig:all} and \ref{fig:rescaling}, we make several important observations.

(i) For all lattices considered, no integrable cases besides the Ising case $\Jm=1$ were found. This indicates that, for infinite lattices of classical spins with the nearest-neighbor interaction, the existence of another integrable case is highly unlikely.

(ii) The value of $\lm$ is mainly controlled by $\Jm$, especially for the bipartite lattices.

(iii) The dependence  $\lm(\Jm)$ for bipartite lattices (L1-L5) has nearly universal form, as illustrated by the rescaling shown in Fig.~\ref{fig:rescaling}.

(iv) The above dependence is nearly flat below $\Jm \approx 0.85$; i.e., away from the integrable limit $\lm$ is very weakly sensitive to the details of microscopic interaction. 

(v) Near the integrable limit $\Jm = 1$, to the best of our numerical accuracy ($1-\Jm > 10^{-8}$), each lattice, bipartite or not, becomes immediately chaotic, and, as shown in Fig.~\ref{fig:all}(b), exhibits an approximate power-law scaling 
$\lm \cong \alpha (1 - \Jm)^{1/3}$, where $\alpha$ is a constant.

(vi) As can be seen from Fig.~\ref{fig:all}(a),
the nonbipartite lattices (L6) and (L7) show
a fork-shaped spread of $\lm$ as $\Jm$ approaches $1/\sqrt{3}$.
The upper and the lower tips of the fork correspond to the anti-Heisenberg and Heisenberg cases, respectively.

Gross features of the above results can be reproduced by rather simple analytical estimates.
In particular, away from the Ising case, the plateau values of $\lm$ seen in Fig.~\ref{fig:all}(a) at $\Jm < 0.85$ can be estimated with a factor-of-two accuracy as the typical frequency of one-spin motion given by the root-mean-squared value of the local field ${\mathbf h}_i$ at the infinite temperature: $\lm \sim  [n (J_x^2 + J_y^2 + J_z^2)/3]^{1/2}$, where $n$ is the number of the nearest neighbors.
This expression also predicts that the heights of these plateaus scale as $n^{1/2}$, while the inset of Fig.~\ref{fig:rescaling} indicates that the actual scaling is close but somewhat steeper.

Now we turn to the approximation for the dependence $\lm(\Jm)$ as the system approaches the Ising limit.
We assume that $J_z \gg J_x, J_y$, which implies that
$\Jm = J_z$. We also introduce variable $J_\perp \equiv [(J^2_x + J^2_y)/2]^{1/2} = [(1-\Jm^2)/2]^{1/2}$ to denote the typical value of the transverse coupling.
We consider two phase space trajectories $\{ {\mathbf S}_i(t) \}$ and $\{ {\mathbf S}_i(t) + \delta {\mathbf S}_i (t) \}$, where $\{ \delta {\mathbf S}_i (t) \}$ is an   infinitesimal difference.
In order to estimate $\lm$,
we linearize the equations of motion~(\ref{eom}) with respect to small $\delta {\mathbf S}_i (t)$ and keep the leading order in terms of $J_{\perp}/J_z$:
\begin{eqnarray}
{d \delta S_{i \varphi} \over dt} &=& J_z \sum_{j(i)} C^{ij}_{z} (t) \delta S_{j z} ~,
\label{d1}
\\
{ d \delta S_{i z} \over dt } &=& J_{\perp} \sum_{j(i)} C^{ij}_{\varphi}(t) \delta S_{j \varphi}~,
\label{d2}
\end{eqnarray}
where $ \delta S_{i z}$ and $ \delta S_{i \varphi}$ are the projection of vectors $ \delta {\mathbf S}_i$ on the directions of vectors ${\mathbf e}_z$ and ${\mathbf e}_z \times {\mathbf S}_i (t)$, respectively.
The third projection of $ \delta {\mathbf S}_i$ does not appear in Eqs.~(\ref{d1}) and~(\ref{d2}), because it can be expressed in terms of $\delta S_{i z}$ -- a consequence of the constraint ${\mathbf S}_i^2 = 1$.
The parameters $C^{ij}_{z}(t)$ and $C^{ij}_{\varphi}(t)$ are determined by 
${\mathbf S}_i (t)$ and ${\mathbf S}_j (t)$ and have characteristic fluctuation times $1/J_{\perp}$ and $1/J_z$, respectively.

Now, we make an assumption justified by the final result [Eq.~(\ref{lm2})]
that $J_{\perp} \ll \lm \ll J_z$.
We estimate the growth of the typical values of $ \delta S_{i z}$ and $ \delta S_{i \varphi}$ over time $\tau$ such that $\lm \ll 1/\tau \ll J_z $.
On the timescale $\tau$, the parameters $C^{ij}_{z}(t)$ stay nearly constant, while 
$C^{ij}_{\varphi}(t)$ strongly fluctuate, so that $\langle C^{ij}_{\varphi}(t) \rangle_{\tau} \approx 0$. We first write
\begin{eqnarray}
\delta S_{i \varphi} (t+\tau ) &\approx & \delta S_{i \varphi} (t) +  \tau J_z \sum_{j(i)} C^{ij}_{z}(t) \delta S_{j z} ~,
\label{D1}
\\
\delta S_{i z} (t+\tau ) &=& \delta S_{i z} (t) +  J_{\perp} \hskip-\smallskipamount \int_{t}^{t+\tau} \! \! \hskip-\bigskipamount dt' \sum_{j(i)} C^{ij}_{\varphi}(t') \delta S_{j \varphi}(t').
\label{D2}
\end{eqnarray}
In this problem, a relatively slow growth of $S_{i \varphi}$ is coupled to a random-walk-like growth of $S_{i z}$. In order to extract $\lm$, we, therefore, look at the leading terms in the growth of $\delta S_{i \varphi}^2$ and $\delta S_{i z}^2$:
\begin{eqnarray}
\delta S_{i \varphi}^2 (t+\tau ) & \! \approx & \! \delta S_{i \varphi}^2 (t) \! + \! 2 \tau J_z 
\!\! \sum_{j(i)} \! C^{ij}_{z}(t) \delta S_{j z}(t) \delta S_{i \varphi} (t),
\label{E1}
\\
\delta S_{i z}^2 (t+\tau ) &=& \delta S_{i z}^2 (t) +  J_{\perp}^2 \int_{t}^{t+\tau} dt'
\int_{t}^{t+\tau} dt^{''}  
\nonumber
\\
&& \hskip-2\bigskipamount\times \sum_{j(i),k(i)} C^{ij}_{\varphi}(t') C^{ik}_{\varphi}(t^{''}) \delta S_{j \varphi}(t') \delta S_{k \varphi}(t^{''}) ~.
\label{E2}
\end{eqnarray}
In Eq.(\ref{E2}), we neglected the term linear in $J_{\perp}$, because $\langle C^{ij}_{\varphi}(t)\rangle_{\tau} \approx 0 $. 

Since we are only concerned with the scaling of $\lm$ with $\Jm$ and $n$, we convert Eqs.~(\ref{E1}) and~(\ref{E2}) into an order-of-magnitude estimate for the typical growth of $\delta S_{i \varphi}^2$ and $\delta S_{i z}^2$.
The estimate includes (i) dropping lattice index in Eqs.~(\ref{E1}) and~(\ref{E2}), (ii) estimating the instantaneous values of parameters $C^{ij}_{z}(t)$ and $C^{ij}_{\varphi}(t)$ by~1, (iii) replacing the sum in Eq.~(\ref{E1}) by $\sqrt{n}$ (a consequence of the random sign of the $n$ terms in that sum), and (iv) approximating the integral term in Eq.~(\ref{E2}) by 
$ J_{\perp}^2 \tau (\delta S_{\varphi})^2  \int_0^{\infty} dt'   \langle \sum_{j(i)} C^{ij}_{\varphi}(t) C^{ij}_{\varphi}(t+t') \rangle_t \sim J_{\perp}^2 \tau (\delta S_{\varphi})^2 \sqrt{n}/J_z $.  The factor $\sqrt{n}/J_z$ in the latter estimate is due to $n$ terms in the sum, each producing a contribution to the integral of the order $1/h_{iz} \sim 1/(J_z \sqrt n)$. Finally, after dividing thus simplified Eqs.~(\ref{E1}) and~(\ref{E2}) by 
$(\delta S_{\varphi})^2$ and $(\delta S_{z})^2$, respectively, we obtain
\begin{eqnarray}
{\delta S_{\varphi}^2 (t+\tau ) \over \delta S_{\varphi}^2 (t)} & \sim & 1 +  2 \tau \sqrt{n} J_z  {\delta S_z(t) \over \delta S_{\varphi} (t)} , 
\label{D1A}
\\
{\delta S_z^2 (t+\tau ) \over \delta S_z^2 (t)} & \sim & 1 +  \tau \sqrt{n} {J_{\perp}^2 \over J_z} { \delta S_{\varphi}^2 (t) \over \delta S_z^2(t)}~.
\label{D2A}
\end{eqnarray}
Since the parameters $\delta S_{\varphi}^2$ and $\delta S_z^2$ should grow at the same rate, this implies that
\begin{equation}
{\delta S_z(t) \over \delta S_{\varphi} (t)}  \cong
\left( { J_{\perp} \over J_z} \right)^{2/3} \approx \left(  1-\Jm \right)^{1/3}.
\label{ratio}
\end{equation}
Substituting Eq.~(\ref{ratio}) into Eq.~(\ref{D2A}) and comparing the right-hand side with the expression $1+ 2 \lm \tau$,
we finally obtain the estimate:
\begin{equation}
\lm \cong
\sqrt{n} \ \Jm^{1/3} \ (1-\Jm^2 )^{1/3} \approx \sqrt{n}  \ (1-\Jm )^{1/3},
\label{lm2}
\end{equation}
which is consistent with the numerically observed 1/3-power law shown in Fig.~\ref{fig:all}(b).
In Ref.~\cite{barnett} the same power law was obtained for weakly interacting dilute gases via a perturbation expansion around the integrable ideal gas.
The prefactor scaling as $\sqrt{n}$ is also compatible with the numerical results, as illustrated in the inset of Fig.~\ref{fig:all}(b).

\begin{figure} \setlength{\unitlength}{0.1cm}

\begin{picture}(100, 55)
{ 
\put(0, 55){ \epsfxsize= 2.3in \rotatebox{270}{\epsfbox{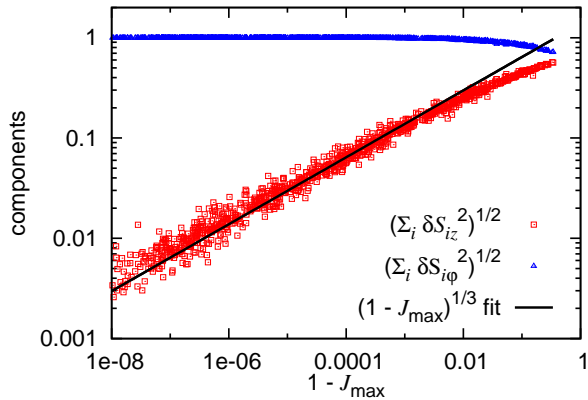}} }
}
\end{picture}
\caption{Projections of the Lyapunov vector correponding to $\lm$ onto the subspaces $\{ \delta S_{iz} \}$ and $\{ \delta S_{i\varphi} \}$  for the $16\times 16 \times 16$ cubic lattice (L5). Solid line is the fit based on Eq.(\ref{ratio}).
}
\label{fig:components} 
\end{figure}

Equation~(\ref{ratio}) predicts further that the components of the Lyapunov eigenvector corresponding to $\lm$ depend systematically on whether they are parallel to the $z$-direction or not. As illustrated in Fig.~\ref{fig:components}, this prediction also agrees with our numerical results.

Our derivation of Eq.~(\ref{lm2}) is based on the assumption of the random fluctuations of the sum in Eq.~(\ref{D2}). However, in the Ising limit, the Fourier transform of this sum contains only a finite number of frequencies. Therefore, in the vicinity of the Ising limit, recurrences may occur if correlations in $S_{iz}$ do not decay sufficiently fast.
Such recurrences, in turn, would contradict our assumption of the fast decay of the time correlations of the above sum.
This would be most problematic for lattices with smaller numbers of nearest neighbors, such as spin chains.
Indeed, we observe in Fig.~\ref{fig:all}(b), that spin chains exhibit larger deviations from the 1/3-power law than other lattices.
The above recurrences may, in fact, be the route to breaking down the chaotic nature of the spin dynamics around the Ising limit.
However, our numerical results indicate that, if such a breakdown occurs, it  occurs at values of $|J_{\perp}/J_z| < 10^{-4}$.

Finally, we mention that the above estimate may be repeated for the case of $J_x$ and $J_y$ smaller but not much smaller than $J_z$.
In this a case, $\tau$ must be chosen much shorter than $1/J_{\perp}$ and thus both $ C^{ij}_{z}(t)$ and $ C^{ij}_{\varphi}(t)$ can be assumed constant on the time scale of $\tau$.
This estimate would then give $\lm \cong (n \Jm)^{1/2} (1-\Jm^2)^{1/4}$, which, as illustrated in Fig.~\ref{fig:rescaling}, exhibits a good overall agreement with the numerical results for bipartite lattices over the entire range of $\Jm$.

In conclusion, we have presented a systematic study of largest Lyapunov exponents $\lm$ for a very large variety of classical spin systems. Our findings strongly suggest the absence of integrable nearest-neighbor Hamiltonians for the type of lattices considered except for the Ising case. As far as the behavior of $\lm$ is concerned, a number of universal features enumerated above as (ii)-(vi) are observed.
We have also analytically derived the scaling of $\lm$ with the anisotropy of the Hamiltonian.

The numerical part of this work was performed at the bwGRiD computing cluster at the University of Heidelberg.
The authors are grateful to T.~A.\ Elsayed and G.~P.\ Morriss for discussions.
A.S.W.'s work is financially supported by an Unga Forskare grant from the Swedish Research Council.


\begin{thebibliography}{10}
\bibitem{footnotespins}%
  \BibitemOpen
  \bibinfo {note} {The equations of motion follow from the Poisson-bracket
  formulation of the Hamiltonian dynamics: $d S_{i,\mu} /dt = \{ {\cal H},
  S_{i,\mu} \}$, where the second index $\mu$ admits values $1$, $2$ or $3$
  representing the spin projections $x$, $y$, or $z$, respectively. The primary
  Poisson brackets in this case are: $ \{ S_{i,\mu}, S_{j,\nu} \} = \delta_{ij}
  \sum_{\kappa} \epsilon_{\mu \nu \kappa} S_{i,\kappa}$, where $\delta_{ij}$ is
  the Kronneker symbol and $\epsilon_{\mu \nu \kappa}$ the Levi-Civita
  symbol.}%
  \bibAnnoteFile{Stop}{footnotespins}%
\bibitem{footnoteinfinite}%
  \BibitemOpen
  \bibinfo {note} {Our ensemble of initial conditions corresponds to 
  infinite temperature in the microcanonical sense. If instead, we were to use
  the canonical infinite-temperature ensemble, this would mean totally random
  sampling in the phase space, which would result in the fluctuations of total
  energy of the order of $1/\sqrt{N_s}$, where $N_s$ is the total number of
  spins. These fluctuations would lead to very small fluctuations of
  $\lambda_{\hbox{max}}$. By setting the total energy to zero we eliminate
  these fluctuations completely, and thus, in particular, guarantee that they
  do not influence our numerical results in the vicinity of the integrable
  Ising limit.}%
  \bibAnnoteFile{Stop}{footnoteinfinite}%
\end{thebibliography}
\end{document}